\documentclass[twocolumn,notitlepage,floatfix,aps,pra,reprint,superscriptaddress,groupedaddress,nofootinbib]{revtex4-2}
\usepackage{graphicx}
\usepackage{ragged2e}
\usepackage[dvipsnames,svgnames]{xcolor}
\usepackage{bm,bbm}
\usepackage{braket}
\usepackage{amssymb,amsmath,amsthm,amsbsy,mathptmx}
\usepackage{amsfonts}
\usepackage{mathtools}
\DeclareMathAlphabet{\mathcal}{OMS}{cmsy}{m}{n}
\newcommand{\dsone}{\text{\usefont{U}{dsrom}{m}{n}1}}
\usepackage{enumerate}
\usepackage{booktabs,multirow}
\usepackage{mathtools,bbding}
\usepackage[percent]{overpic}
\usepackage{microtype}
\usepackage{array,adjustbox,afterpage}
\usepackage[T1]{fontenc}
\usepackage[utf8]{inputenc}
\usepackage{tikz}
\usetikzlibrary{calc}
\usepackage[english]{babel}
\usepackage{newtxtext,newtxmath}

\makeatletter
\renewcommand{\fnum@figure}{FIG. \thefigure} 
\makeatother

\usepackage[bookmarks=false,colorlinks=true,linkcolor=blue,citecolor=purple,urlcolor=purple]{hyperref}
\usepackage{cleveref}
\usepackage{subfloat}
\usepackage[normalem]{ulem}

\Crefname{subfigures}{figure}{figures}
\Crefname{subfigures}{Figure}{Figures}

\usepackage{pifont}

\def\bra#1{\mathinner{\langle{#1}|}}
\def\ket#1{\mathinner{|{#1}\rangle}}
\def\braket#1#2{\langle{#1}|#2\rangle}
\def\ketbra#1#2{|{#1}\rangle\langle{#2}|}

\hypersetup{
    pdfstartview={FitH} 
    }

\begin{document}


\title{Coherence, superposition, and L\"{o}wdin symmetric orthogonalization}

\author{G\"{o}khan Torun}
\email{gtorung@gmail.com}
\affiliation{TÜBİTAK Research Institute for Fundamental Sciences, 41470 Gebze, T\"{u}rkiye}

\begin{abstract}
The notions of coherence and superposition are conceptually the same; however, an important distinction exists between their resource-theoretic formulations. Namely, while basis states are orthogonal in the resource theory of coherence, they are not necessarily orthogonal in the resource theory of superposition. Owing to the nonorthogonality, the manipulation and characterization of superposition states require significant efforts. Here, we demonstrate that the L\"{o}wdin symmetric orthogonalization (LSO) method offers a useful means for characterizing pure superposition states. The principal property of LSO is that the structure and symmetry of the original nonorthogonal basis states are preserved to the greatest extent possible, which prompts us to study the role of LSO in identifying the hierarchical relations of resource states. Notably, we reveal that the maximally coherent states turn into the states with maximal superposition with the help of LSO: in other words, they are equivalent under the action of symmetric orthogonalization. Our results facilitate further connections between coherence and superposition, where LSO is the main tool.
\end{abstract}

\maketitle



\section{Introduction}\label{Sec:Intro}

Characterizing the intrinsic physical features of quantum mechanics, which is what provides the way to understand their pivotal role as a resource in information processing tasks, is among the essential topics in quantum information science. In this regard, the study of quantum resource theories \cite{Horodecki-QRT, Brandao2015, COECKE201659, Chitambar-QRTs, Regula-QRTs} has gained prominence as an active area of research over time as well. A fundamental concept with particular relevance is that of quantum coherence \cite{Aberg-Superposition, Baumgratz-Coherence} which lies at the heart of many quantum information processing applications, including quantum metrology \cite{Iman2016Speakable, piani_robustness_2016, DegenQSensing2017, Giovannetti2011}, quantum computation \cite{Hillery2016QC}, and quantum thermodynamics \cite{Brandao2013TEq, GOUR20151, Lastaglio2015COH, Goold_2016, Pusuluk2021TCoh}.
Basically, the resource theory of coherence (RTC) \cite{Baumgratz-Coherence} is formulated with respect to a reference basis that forms an orthonormal set. In addition, a resource-theoretic formulation based on nonorthogonal basis states generalizes the concept of coherence, known as the resource theory of superposition (RTS) \cite{Plenio-RTofS}, where a wide range of tasks such as communication complexity \cite{Feix-CComp, Brukner-Complexity} and channel discrimination \cite{Plenio-RTofS} need the presence of superposition as a resource. Thus, from the perspective of resource theory, coherence is accepted as a special case of superposition, i.e., in the limit of overlapping between basis states going to zero, RTS reduces to RTC. Comparing the amount of studies based on RTC and RTS, the former has made significant progress \cite{Streltsov-CoherenceRT}, whereas the latter needs further research. Nonorthogonality is a major factor that makes it difficult to contribute to this need. As a consequence, it is natural and necessary to delve deeper into the links between RTC and RTS, which enhances our capabilities in exploring and exploiting the latter.

It is our purpose in this paper to present a systematic approach to the characterization of superposition states by using L\"{o}wdin symmetric orthogonalization (LSO) \cite{Lowdin1950, Aiken1980, Mayer2002, PIELA2014e99} as an instrumental tool. To the best of our knowledge, our study is the first to shed light on the aforementioned problem in this way, which offers a great potential to extend the results obtained for the coherence to the superposition. A number of studies have included LSO, such as \cite{Horoshko2017SCats} which considered the Schmidt decomposition of the two-mode Schrödinger cat states --- a superposition of two distinct coherent states --- and \cite{WALK2012649} whose objective was to obtain a set of orthogonal (L\"{o}wdin) pulse. Among the different types of orthogonalization schemes \cite{Srivastava2000}, our main object of interest is the LSO where the importance of this method is due to the fact that it preserves as much as possible of the structure and symmetry of the original nonorthogonal basis \cite{Lowdin1950, PIELA2014e99}. For instance, this attractive feature is the main reason why LSO is preferred when one aims to obtain a set of orthonormal orbitals generated from the Hilbert space spanned by the nonorthogonal orbitals \cite{Pratt1956MOO, Carlson1957OrtPro}. These insights allow us to probe more deeply into the connections between the coherence and the superposition --- providing a picture of the utility of LSO for characterizing and manipulating the superposition states.

The task of state transformation, as a central subject of study within resource theories, is concerned with converting a given state into a target state, probabilistic or deterministic, via quantum operations that do not create resource. The literature on RTC comprises a diverse range of studies that investigate the manipulation of coherence through various techniques \cite{Aberg_CC, Du-CoherenceMaj, Xiao2015IRMQC, Winter2016ORTofC, Korzekwa_2016, de_Vicente_2016, Kaifeng_CC, Streltsov2017Structure, 2018Regula-OneShotCohDis, Torun-DetCoherence, Zhao2018Dilution, Regula2018Nonasymptotic, Kun2018PDQC, 2019Liu-DetCoh, Lami2019GenericBound, Torun-CoherenceDistill, Regula-OneShotDistill, Regula_DepCovOper, Liu2022Distill, Liu2021OptCoherenceDistill, Liu2022ApprDistillCoh, Regula2022PT, Takagi2022Catalysts}. In the case of RTS, however, a partial order structure was uncovered for the probabilistic transformations of pure superposition states \cite{Plenio-RTofS}. Later on, deterministic transformations of superposition states was investigated in Ref.~\cite{torun2020resource}. Of course, it is always essential to consider a well-established method that offers an alternative and practical view for the state transitions between pairs of quantum states, which also enables a broad characterization of resource states. Importantly, from the perspective of characterizing states, a state at the pinnacle of the resourcefulness hierarchy often considered a valuable unit of resource. Such (maximally resourceful) states are well defined in the resource theory of (bipartite) entanglement, \(\ket{\Psi_{d}}=({1}/{\sqrt{d}}) \sum_{i=1}^{d}\ket{ii}\) \cite{Horodecki-QE} (see \cite{Tejada2019MRE} for multipartite systems); coherence, \(\ket{\Psi_{d}}=({1}/{\sqrt{d}}) \sum_{i=1}^{d}\ket{i}\) \cite{Baumgratz-Coherence, Peng2016MCS}; and imaginarity, \(\ket{\hat{\mp}}=(\ket{0} \mp i\ket{1})/\sqrt{2}\) \cite{Hickey_2018, Wu2021RTofImaginarity}.

Central to our study is the characterization of the states with maximal superposition, with the potential to illuminate the transformation of superposition states. The recent work of Ref.~\cite{Senyasa2022Golden} has established a structural way to investigate maximal states of RTS, i.e., golden states, utilizing the eigenvalues and eigenvectors of the Gram matrix constructed from the given set of arbitrary (nonorthogonal) basis states. Here we show that it is possible to effectively accomplish the same goal, where LSO allows us to do so. In undertaking efforts to achieve this goal, we proceed exactly as follows (see Fig.~~\ref{Fig2}). We are given a quantum state \(\ket{\tau}\), which can be represented by either the state \(\ket{\psi}\) or the state \(\ket{\bar{\psi}}\), both of which can be converted into each other using LSO. Specifically, while the state \(\ket{\tau}\) describes a superposition state, \(\ket{\psi}\), in nonorthogonal basis, it describes a coherent state, \(\ket{\bar{\psi}}\), in orthogonal basis. Then, when the state \(\ket{\bar{\psi}}\) is maximally coherent, the question presents itself: Which state does \(\ket{\psi}\) correspond to in the case of superposition? Our analysis shows that in this scenario, \(\ket{\psi}\) corresponds to the state with maximal superposition. In other words, the pair of states \(\ket{\psi}\) and \(\ket{\bar{\psi}}\) (both represents the same physical state \(\ket{\tau}\)) are equivalent under LSO, with \(\ket{\psi}\) being the state with maximal superposition and \(\ket{\bar{\psi}}\) being the maximal coherent state. We believe that LSO, a well-known scheme in quantum chemistry that serves as the enabling tool for the analysis described above, has the potential to advance the state of the art in our understanding of the concepts of coherence and superposition within the scope of resource theory.

The remainder of this paper is structured as follows. In Sec.~\ref{Sec:Definitions}, we give the background that contains the necessary ingredients and tools needed to construct the paper: the basics of RTS and RTC, and the details of LSO. We proceed in Sec.~\ref{Sec:Results} by discussing the role of LSO in the characterization of coherence and superposition (the main focus is on superposition). In Sec.~\ref{SubSec:MS}, we employ LSO to reveal the form of the states with maximal superposition and present our main results. We conclude our work in Sec.~\ref{Sec:Conc}.


\section{Background}\label{Sec:Definitions}

\subsection{Resource theory of superposition}\label{SubSec:RTS}

The essentials of RTS was introduced and discussed in Ref.~\cite{Plenio-RTofS}. Here we give a succinct overview of RTS, including the free states (\(\mathcal{F}\)), resource states (\(\mathcal{R}\)), and the restricted set of free operations (\(\mathcal{O}\)), which are the three major elements of a general quantum resource theory \cite{Chitambar-QRTs}.
As in \cite{Plenio-RTofS}, let \(\mathcal{H}_d\) be a \(d\)-dimensional Hilbert space and \(\{\ket{c_i} : i=0,1,\dots,d-1\}\) be the set of nonorthogonal, normalized, and linearly independent basis of \(\mathcal{H}_d\).
Then, states defined as \(\rho = \sum_i \rho_i \ketbra{c_i}{c_i}\) are called superposition-free, where \(\rho_i \geq 0\) form a probability distribution. The set of superposition-free density operators is denoted by \(\mathcal{F}\) and forms the set of free states. All density operators which are not an element of \(\mathcal{F}\) are called
superposition states and form the set of resource states \(\mathcal{R}\).
A linear combination of \(\{\ket{c_i}\}\) gives us the pure superposition states
\begin{eqnarray}
\ket{\psi} = \sum_{i=0}^{d-1} \psi_{i} \ket{c_i},
\end{eqnarray}
where the coefficients \(\{\psi_i\}\) are complex number. Let \(S\) be overlap matrix of basis states\(\{\ket{c_i}\}\), i.e., Gram matrix,
with the elements \(S_{ij}=\braket{c_i}{c_j}\). Then the normalization condition reads \(\braket{\psi}{\psi}=\sum_{i,j=0}^{d-1} \psi_{i}^{\ast} S_{ij} \psi_{j}=1\), where \(S_{ij}\) are complex in general.

In the context of RTS, a Kraus operator \(K_n\) is called superposition-free if \(K_n \rho K_n^{\dagger} \in \mathcal{F}\)
for all \(\rho \in \mathcal{F}\), and is of the form
\begin{eqnarray}\label{Kraus-SupFree}
K_n=\sum_{k} c_{k,n} {\ket{c_{f_{n}(k)}}\bra{c_k^{\perp}}},
\end{eqnarray}
where \(c_{k,n}\) are complex number, \({f_{n}(k)}\) are arbitrary index functions~\cite{Plenio-RTofS}, and \(\braket{c_i^{\perp}}{c_j}=\zeta_i\delta_{ij}\) for \(\zeta_i\in \mathbb{C}\) where the vectors \(\ket{c_k^{\perp}}\) are normalized.
Also, a quantum operation \(\Phi(\cdot)\) is called superposition-free if it is trace preserving and can be written
such that \(\Phi(\rho) = \sum_{n} K_n \rho K_n^{\dagger}\), where all \(K_n\) are free. For further reading and details, we refer the reader to Refs.~\cite{Plenio-RTofS, torun2020resource, Senyasa2022Golden}.

\subsection{L\"{o}wdin symmetric orthogonalization}\label{SubSec:LowdinSO}

Recall that \(\{\ket{c_i} : i=0,1,\dots,d-1\}\) be a set of nonorthogonal, normalized, and linearly independent basis vectors as in the previous section. We aim to construct an orthonormal basis set \(\{\ket{l_i} : i=0,1,\dots,d-1\}\) by a suitable linear transformation.
To this end, we utilize the L\"{o}wdin's symmetric orthogonalization \cite{Lowdin1950,PIELA2014e99} abbreviated by LSO, so that the orthogonal basis vectors \(\{\ket{l_i}\}\) are called as ``L\"{o}wdin basis''. We start by defining column vectors
\begin{equation}\label{vector-forms}
\bm{C} := {\left(c_0, c_1, \dots, c_{d-1} \right)}^\intercal, \quad
\bm{L} := {\left(l_0, l_1, \dots, l_{d-1} \right)}^\intercal.
\end{equation}
Here the components \(\{c_i\}\) and \(\{l_i\}\) represent the basis sets \(\{\ket{c_i}\}\) and \(\{\ket{l_i}\}\), respectively.
We can define a general linear transformation \(T\) such that
\begin{eqnarray}\label{LowdinTransformation}
\bm{L} = T \bm{C}, \quad \Big(TT^{\dagger}=S^{-1}\Big).
\end{eqnarray}
In order to get the linear transformation \(T\), we first construct the overlap matrix \(S\), which is positive semidefinite in general, such that \(S_{ij}=\braket{c_i}{c_j}\) for \(i,j=0,1,\dots,d-1\). By a unitary matrix \(U\), we can diagonalize the overlap matrix \(S\), that is, \(S_{\text{diag}} = U^{\dagger} S U\). Let \(S\) have eigenvalues \(\{\lambda_i\}\). Since the eigenvalues of \(S\) are positive, the elements of \(S_{\text{diag}}\) can be replaced by \(\{\lambda_i\}\), from which we can get \({S^{{1}/{2}}_{\text{diag}}}=\text{diag}(\sqrt{\lambda_0}, \sqrt{\lambda_1}, \dots, \sqrt{\lambda_{d-1}})\). By using this matrix, we define the matrix \({S^{{1}/{2}}}=U S^{{1}/{2}}_{\text{diag}}U^{\dagger}\) and finally get \({S^{-{1}/{2}}}={(S^{{1}/{2}})}^{-1}\). Then the linear transformation given by Eq.~\eqref{LowdinTransformation} can be rewritten as
\begin{eqnarray}\label{LSO-Elements}
\big[\bm{L}\big]_{i} = \sum_{j=0}^{d-1} \big[{S^{-{1}/{2}}}\big]_{ij} \big[\bm{C}\big]_{j},
\end{eqnarray}
for \(i,j=0,1,\dots,d-1\), which is invertible. Needless to say, here \([{S^{-{1}/{2}}}]_{ij}\) is the (\(i, j\))-th matrix element of \({S^{-{1}/{2}}}\) and \([\bm{L}]_{i}\) (\([\bm{C}]_{i}\)) is the \(i\)-th entry of \(\bm{L}\) (\(\bm{C}\)). Thus, we can obtain the orthogonal basis vectors \(\{\ket{l_i}\}\) as a linear combination of the nonorthogonal basis vectors \(\{\ket{c_i}\}\), that is, from Eq.~\eqref{LSO-Elements} one has \(\ket{l_i}=\sum_{j}[{S^{-{1}/{2}}}]_{ij}\ket{c_j}\).


\begin{figure}[t]
	\centering
	\includegraphics[width=1\columnwidth]{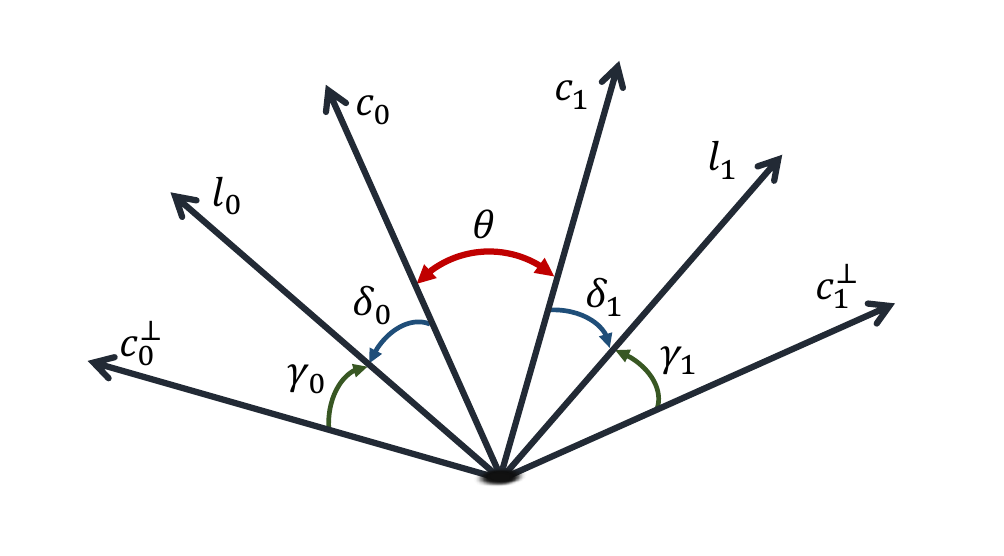}
	\caption{A geometric visualization for two linearly independent nonorthogonal vectors \({c_0}\) and \({c_1}\) with reciprocal states \({c_0^{\perp}}\) and \({c_1^{\perp}}\) \cite{Plenio-RTofS}. The basis sets \(\{c_0, c_1\}\) and \(\{c_0^{\perp}, c_1^{\perp}\}\) can be expressed in terms of symmetrically orthonormalized vectors \({l_0}\) and \({l_1}\). In order to satisfy Eq.~\eqref{LSO-min} one needs to take \(\delta_0=\delta_1=\delta\), and thus one has \(\theta+2\delta={\pi}/{2}\). For normalized vectors \({c_0}\) and \({c_1}\), the inner product (i.e., overlap) reads \(\braket{c_0}{c_1}=\cos \theta\) with \(\theta \in (0,\pi)\). Similarly, for normalized vectors \({c_0^{\perp}}\) and \({c_1^{\perp}}\), one has \(\gamma_0=\gamma_1=\delta\) and \(\braket{c_0^{\perp}}{c_1^{\perp}}=-\cos \theta\). Also, here \(\braket{c_i}{c_i^{\perp}}=\sin \theta\) and \(\braket{c_i}{c_j^{\perp}}=0\) for \(i,j=0,1\) (\(i \neq j\)). Thus, the symmetric orthonormalization procedure generates new vectors \({l_0}\) and \({l_1}\) which are symmetrically rotated respect to \({c_0}\) and \({c_1}\).}
	\label{Fig1}
\end{figure}

In fact, the ways of taking a nonorthogonal set into an orthonormal set are multifarious. The LSO has a meaningful feature among all possible orthogonalization methods: namely, the symmetric orthogonalization ensures
\begin{eqnarray}\label{LSO-min}
\sum_{i} \|c_i - l_i\|^2=\min,
\end{eqnarray}
where \(\|c_i - l_i\|^2 \equiv \braket{c_i - l_i}{c_i - l_i}\). From Eq.~\eqref{LSO-min}, one can infer that the symmetrically orthogonalized vectors \(\{\ket{l_i}\}\) are the least distant in the Hilbert space from the original nonorthogonal vectors \(\{\ket{c_i}\}\), i.e., LSO implies the gentlest pushing of the directions of the nonorthogonal vectors in order to get them orthogonal \cite{Lowdin1950,PIELA2014e99}.

One can consider two nonorthogonal states \(\ket{c_0}\) and \(\ket{c_1}\), as in Fig.~\ref{Fig1}, to illustrate this point further. As sketched in Fig.~\ref{Fig1}, by symmetrically rotating nonorthogonal states \(\ket{c_0}\) and \(\ket{c_1}\), then orthogonal states \(\ket{l_0}\) and \(\ket{l_1}\) are generated such that \(\|c_0 - l_0\|^2+\|c_1 - l_1\|^2=\min\). Moreover, another two nonorthogonal basis states \(\ket{c_0^{\perp}}\) and \(\ket{c_1^{\perp}}\) --- reciprocal states --- also exist which take part in the definition of superposition-free operations for qubit systems. Thus, it is readily apparent that the three basis sets \(\{\ket{c_0}, \ket{c_1}\}\),  \(\{\ket{c_0^{\perp}}, \ket{c_1^{\perp}}\}\), and \(\{\ket{l_0}, \ket{l_1}\}\) can be expressed in terms of each other, where LSO brings out these connections in the sense of Eqs.~\eqref{LSO-Elements} and \eqref{LSO-min}. Therefore, LSO may provide a comprehensive understanding of the relationship between the RTC and RTS frameworks.

\subsection{Resource theory of coherence}\label{SubSec:RTC}

The resource theory of coherence (RTC), which is a special case of RTS, is a basis dependent concept. Importantly, one may choose the reference basis in accordance with the physics of the problem under consideration. In that connection, we here designate the L\"{o}wdin basis as our preferred (complete and orthonormal) basis. Namely, let \(\{\ket{l_k}: k=0,1,\dots,d-1\}\) be a particular orthonormal basis of the \(d\)-dimensional Hilbert space \(\mathcal{H}_d\); then the quantum states that are diagonal with respect to this specific basis \(\{\ket{l_k}\}\) are called as incoherent, constituting a set labeled by \(\mathcal{I}\) \cite{Aberg-Superposition,Baumgratz-Coherence}. Hence, all incoherent states \(\rho \in \mathcal{I}\) are of the form
\(\rho=\sum_{k=0}^{d-1} p_k \ket{l_k}\bra{l_k}\), where \(p_k\in [0,1]\) and \(\sum_{k}p_k=1\). On the other hand, a finite $d$-dimensional pure coherent state is given by
\begin{eqnarray}\label{CoherentStatesGenForm}
\ket{\phi}=\sum_{k=0}^{d-1} \phi_k e^{i\theta_k} \ket{l_k},
\end{eqnarray}
where \(\{\phi_k: k=0,1,\dots,d-1\}\) are real numbers such that \(\phi_k \geq \phi_{k+1}> 0\), satisfying \(\sum_k\phi_k^2=1\).
Note that all complex phases \(\theta_k \in [0,2\pi)\) in Eq. \eqref{CoherentStatesGenForm} can be eliminated by incoherent operations.
We denote by \(\ket{\Phi_d}\) the maximally coherent state in the reference basis with entries \(\phi_k=\frac{1}{\sqrt{d}}\),
that is,
\begin{eqnarray}\label{MaximallyCoherentState}
\ket{\Phi_d}=\frac{1}{\sqrt{d}}\sum_{k=0}^{d-1} e^{i\theta_k} \ket{l_k}.
\end{eqnarray}
Moreover, since all other \(d\)-dimensional coherent states can be generated from \(\ket{\Phi_d}\) by means of the free operations, this definition regards as a unit of RTC --- coherence bit.

Following standard definitions of RTC, Kraus operators \(\{K_n\}\) are called incoherent operators,
which are characterized as the set of trace preserving completely positive maps, such that
\begin{eqnarray}\label{FreeOperCoh}
\rho \rightarrow \rho_n=\frac{K_n\rho K_n^{\dag}}{\text{Tr}[K_n\rho K_n^{\dag}]} \in \mathcal{I}
\end{eqnarray}
for all \(n\), where \(\rho \in \mathcal{I}\) and \(\sum_{n}K_n^{\dag}K_n = \dsone\). Also, the quantum operations \(\Lambda(\cdot)\) consisting of the Kraus operators \(\{K_i\}\) satisfying Eq. \eqref{FreeOperCoh} are the free operations in the context of RTC.

\subsection{Quantifying superposition and coherence}\label{SubSec:QuantifyingResources}

At this point, we give two of superposition (and coherence) measures, where
the conditions when a function \(M\) is called a superposition measure were detailed in Refs.~\cite{Aberg-Superposition,Plenio-RTofS}.
In fact, the following two measures of superposition were introduced \cite{Plenio-RTofS} by extending the method used for quantifying coherence \cite{Baumgratz-Coherence}.
Within certain types of resource measures, the \(l_1\) measure of superposition is defined as
\begin{eqnarray}\label{L1MeasureSuperposition}
M_{l_1}(\rho)=\sum_{i \neq j} |\rho_{ij}|,
\end{eqnarray}
for \(\rho=\sum_{i,j} \rho_{ij}\ket{c_i}\bra{c_j}\) \cite{Plenio-RTofS}. Also, Eq.~\eqref{L1MeasureSuperposition} corresponds to the \(l_1\) norm of coherence such that \(\rho=\sum_{i,j} \rho_{ij}\ket{l_i}\bra{l_j}\) \cite{Baumgratz-Coherence}.
The relative entropy of superposition is given by
\begin{eqnarray}\label{RelEntSuperposition}
M_{\text{rel.ent}}(\rho)=\min_{\sigma \in \mathcal{F}} S(\rho \| \sigma),
\end{eqnarray}
where \(S(\rho \| \sigma)=\text{tr}[\rho \log \rho]-\text{tr}[\rho \log \sigma]\) denotes the quantum relative entropy.
From the definition of \eqref{RelEntSuperposition}, the relative entropy of coherence reads \(S(\rho_{\text{diag}})-S(\rho)\), where incoherent state \(\rho_{\text{diag}}\) denotes the state obtained from \(\rho\) by deleting all off-diagonal elements \cite{Baumgratz-Coherence}.


\section{Setting the Scene: Two Representations of a Given Quantum state}\label{Sec:Results}

The purpose of this section is to explicitly construct two different representations of a given pure quantum state such that one of the two representations involves a nonorthogonal basis consisting of vectors \(\{\ket{c_0}, \dots \ket{c_{d-1}}\}\) and the other involves an orthogonal basis consisting of vectors \(\{\ket{l_0}, \dots, \ket{l_{d-1}}\}\), where LSO is the applied technique that relates these to each other.

\subsection{Qubit systems}\label{SubSec:Qubits}

To formulate our approach based on LSO, it is natural to start with exploring the two-dimensional systems. Let us then consider the nonorthogonal basis states \(\ket{c_0}\) and \(\ket{c_1}\) with \(\braket{c_0}{c_1}=s\in (-1,1)\). After concluding the symmetric orthogonalization given by Eq.~\eqref{LSO-Elements}, we obtain
\begin{eqnarray}\label{2D-LSO}
\ket{l_0}=\alpha \ket{c_0} + \beta \ket{c_1}, \quad \ket{l_1}=\beta \ket{c_0} + \alpha \ket{c_1},
\end{eqnarray}
where \(\alpha=({1}/{\sqrt{\lambda_1}}+{1}/{\sqrt{\lambda_0}})/2\) and \(\beta=({1}/{\sqrt{\lambda_1}}-{1}/{\sqrt{\lambda_0}})/2\).
Here, \(\lambda_0={1-s}\) and \(\lambda_{1}={1+s}\) are the eigenvalues of the overlap matrix \(S\).
Thus, from Eq.~\eqref{2D-LSO}, we can complete the linear map \(\{\ket{c_0}, \ket{c_1}\} \overset{\text{LSO}^{\mapsto}}{\longrightarrow} \{\ket{l_0}, \ket{l_1}\}\). Notice that Löwdin basis states \(\ket{l_0}\) and \(\ket{l_1}\) given in Eq.~\eqref{2D-LSO} are superposition states, where their \(l_1\) measures of superposition are equal, \(M_{l_1}(\ket{l_0})=M_{l_1}(\ket{l_1})=|{s}/{(1-s^2)}|\). Also, from Eq.~\eqref{2D-LSO} it is easy to obtain
\begin{eqnarray}\label{2D-LSOInverse}
\ket{c_0}=\alpha'\ket{l_0}-\beta'\ket{l_1}, \quad \ket{c_1}=\alpha'\ket{l_1}-\beta'\ket{l_0},
\end{eqnarray}
where \(\alpha'=\sqrt{1-s^2}\alpha\) and \(\beta'=\sqrt{1-s^2}\beta\). Again, notice that the nonorthogonal basis states \(\ket{c_0}\) and \(\ket{c_1}\) (which are pure superposition-free states as well \cite{Plenio-RTofS}) given in Eq.~\eqref{2D-LSOInverse} are coherent states in terms of Löwdin basis, and their \(l_1\) measures of coherence are equal, \(M_{l_1}(\ket{c_0})=M_{l_1}(\ket{c_1})=|{s}|\).

Now consider a given resource state \(\ket{\tau_2}\) for which the (normalized) superposition state \(\ket{\psi}=\psi_0\ket{c_0}+\psi_1\ket{c_1}\) with \(\braket{c_0}{c_1}=s\in (-1,1)\) is a representation with respect to the nonorthogonal basis \(\{\ket{c_0}, \ket{c_1}\}\). With the help of Eq.~\eqref{2D-LSOInverse}, the conversion \(\ket{\psi} \overset{\text{LSO}^{\mapsto}}{\longrightarrow} \ket{\bar{\psi}}\) can be attained such that
\begin{eqnarray}\label{QubitLSOformRight-Initial}
\ket{\psi} \overset{\text{LSO}^{\mapsto}}{\longrightarrow} \ket{\bar{\psi}} &=&
\big(\alpha'\psi_0-\beta'\psi_1\big)\ket{l_0}
+\big(\alpha'\psi_1-\beta'\psi_0\big)\ket{l_1}
\nonumber \\
&=& {\bar{\psi}}_0 \ket{l_0} + {\bar{\psi}}_1 \ket{l_1}.
\end{eqnarray}
Here in Eq.~\eqref{QubitLSOformRight-Initial} one can see that \({\bar{\psi}}_0\) (\({\bar{\psi}}_1\)) is equal to the conjugate of \({\bar{\psi}}_1\) (\({\bar{\psi}}_0\)) satisfying \(|{\bar{\psi}}_0|^2+|{\bar{\psi}}_1|^2=1\). Thus, we have a superposition state \(\ket{\psi}\) and after applying \(\text{LSO}^{\mapsto}\) we have a coherent state \(\ket{\bar{\psi}}\) in our hand, both representing the same quantum state \(\ket{\tau_2}\). Moreover, the conversion \(\ket{\bar{\psi}} \overset{\text{LSO}^{\mapsfrom}}{\longrightarrow} \ket{\psi}\) can be realized with the help of Eq.~\eqref{2D-LSO}, that is,
\begin{eqnarray}\label{QubitLSO-Inverse}
\ket{\bar{\psi}}  \overset{\text{LSO}^{\mapsfrom}}{\longrightarrow} \ket{\psi}  &=&
\big(\alpha{\bar{\psi}}_0+\beta{\bar{\psi}}_1\big)\ket{c_0}
+\big(\alpha{\bar{\psi}}_1+\beta{\bar{\psi}}_0\big)\ket{c_1}
\nonumber \\
&=& {\psi}_0 \ket{c_0} + {\psi}_1 \ket{c_1}.
\end{eqnarray}
Notice that, while the linear map \(\text{LSO}^{\mapsto}\) seen in Eq.~\eqref{QubitLSOformRight-Initial} corresponds the LSO itself, the linear map \(\text{LSO}^{\mapsfrom}\) here in Eq.~\eqref{QubitLSO-Inverse} corresponds to the LSO, but in the opposite direction.
As a consequence, two pure resource states \(\ket{\psi}\) and \(\ket{\bar{\psi}}\) linked in this way represent the same resource state \(\ket{\tau_2}\). Namely, the former written in terms of nonorthogonal basis, and therefore describes a superposition state; the latter written in terms of Löwdin basis instead of nonorthogonal basis, and therefore describes a coherent state:
\begin{equation}\label{RSConnection}
    \ket{\tau_2} \equiv
    \begin{cases}
      \ket{\psi}: & \text{with respect to}  \,  \{\ket{c_0}, \ket{c_1}\}, \\
      \ket{\bar{\psi}}: & \text{with respect to} \,  \{\ket{l_0}, \ket{l_1}\},
    \end{cases}
  \end{equation}
which we write \(\ket{\psi} \dashleftarrow \ket{\tau_2} \dashrightarrow \ket{\bar{\psi}}\) in short. In addition to this, \(l_1\) measure of superposition of \(\ket{\psi}\) and \(l_1\) measure of coherence of \(\ket{\bar{\psi}}\) are
\begin{eqnarray}\label{L1MeasureSupCoh}
M_{l_1}(\ket{\psi})=2|\psi_0\psi_1|, \quad
M_{l_1}(\ket{\bar{\psi}})=2|\psi_0\psi_1(1-s^2)+\frac{s}{2}|, \quad
\end{eqnarray}
where \(M_{l_1}(\ket{\psi})=M_{l_1}(\ket{\bar{\psi}})\) for \(s=0\) as expected.
We conclude this consideration for qubit systems here by stating that we address the practical advantages of this connection, in Section \ref{SubSec:MS}, simply expressed as given by Eq.~\eqref{RSConnection}.

\subsection{Qudit systems}\label{SubSec:DimensionD}

Having introduced our strategy for two-dimensional systems, it is now time to look at \(d\)-dimensional systems. In this general case, Eq.~\eqref{QubitLSOformRight-Initial} can be rewritten as
\begin{eqnarray}\label{QubitLSOformRight-InitialGeneral}
\ket{\psi}= \sum_{j=0}^{d-1} \psi_j \ket{c_j} \overset{\text{LSO}^{\mapsto}}{\longrightarrow} \ket{\bar{\psi}} =
\sum_{i=0}^{d-1} {\bar{\psi}}_i \ket{l_i}.
\end{eqnarray}
Here the coefficients  \({\bar{\psi}}_i\) are functions of \(\psi_j\) and \(\lambda_k\),
that is, \({\bar{\psi}}_i={\bar{\psi}}_i(\psi_0, \psi_1, \dots, \psi_{d-1}, \lambda_0, \lambda_{d-1})\),
satisfying \(\sum_{i=0}^{d-1} |{\bar{\psi}}_i|^2=1\), where \(\{\lambda_k\}\) are the eigenvalues of the overlap matrix \(S\). As a result,
LSO given in Eq.~\eqref{QubitLSOformRight-InitialGeneral} leads the superposition state \(\ket{\psi}\) to the coherent state \(\ket{\bar{\psi}}\), basically \(\ket{\psi} \dashleftarrow \ket{\tau_d} \dashrightarrow \ket{\bar{\psi}}\).
Given the overlap matrix \(S\) and the superposition state \(\ket{\psi}\) for dimension \(d\), the explicit forms of the coefficients \({\bar{\psi}}_i\) follow by Eq.~\eqref{LSO-Elements}.

For the simplicity of the discussion, we limit ourselves with the inner product setting \(\braket{c_i}{c_j}=s \in \mathbb{R}\). Given the linear independence of basis states, the interval \((\frac{1}{1-d},1)\) encompasses the allowable values for \(s\).  This range signifies that \(s\) lies between \(\frac{1}{1-d}\) and \(1\), exclusively. Consequently, the elements of the overlap matrix, \(S_{ij}=\braket{c_i}{c_j}\), are determined to be real values represented by \(s\). In such an instance, the eigenvalues of the constructed overlap matrix \(S\) are found to be \(\lambda_0=\lambda_1=\dots=\lambda_{d-2}={1-s}\) and \(\lambda_{d-1}={1+(d-1)s}\). This result highlights the equality of eigenvalues \(\lambda_{0}\) to \(\lambda_{d-2}\), each equivalent to \(1-s\), while the eigenvalue \(\lambda_{d-1}\) is determined as \(1+(d-1)s\).

Now, the step-by-step derivation of the matrix \(S^{-1/2}\) (i.e., the transformation matrix \(T\)) is presented herein, demonstrating the systematic process of obtaining the matrix \(S^{-1/2}\) for any given \(d \geq 2\) level as well. The overlap matrix \(S\) can be diagonalized by a unitary matrix \(U\), yielding the diagonal matrix \(S_{\text{diag}} = U^{\dagger} S U\). Notably, the columns of the matrix \(U\) correspond to the eigenvectors of the overlap matrix. This diagonalization process, facilitated by the unitary transformation, enables us to express the overlap matrix \(S\) in a simplified and more interpretable form, with the eigenvectors serving as the basis for this transformation. Since the eigenvalues of the overlap matrix \(S\) are positively defined, it is feasible to replace the elements in \(S_{\text{diag}}\) with the corresponding eigenvalues, denoted as \(\left\{\lambda_i\right\}\). Thus, by taking the square root of each eigenvalue, the resulting matrix \(S^{1/2}_{\text{diag}}\) can be obtained as \(\text{diag}(\sqrt{\lambda_0}, \sqrt{\lambda_1}, \dots, \sqrt{\lambda_{d-1}})\). After that, utilizing the matrix \(S^{1/2}_{\text{diag}}\), we proceed to establish the matrix \(S^{1/2}\) through the expression \({S^{{1}/{2}}}=U S^{{1}/{2}}_{\text{diag}}U^{\dagger}\). Subsequently, we ultimately arrive at the matrix \({S^{{-1}/{2}}}\) by computing the inverse of \({S^{{1}/{2}}}\) denoted as \({\left(S^{{1}/{2}}\right)}^{-1}\). Finally, setting
\begin{eqnarray}\label{TheTermMu}
\mu:=\frac{1}{d}\left[\frac{1}{\sqrt{1+(d-1)s}}+\frac{d-1}{\sqrt{1-s}}\right]
\end{eqnarray}
and
\begin{eqnarray}\label{TheTermKappa}
\kappa:=\frac{1}{d}\left[\frac{1}{\sqrt{1+(d-1)s}}-\frac{1}{\sqrt{1-s}}\right],
\end{eqnarray}
the \(d \times d\) transformation matrix \(S^{-1/2}\) can be brought to the following form:
\begin{eqnarray}\label{OperationT}
    S^{-1/2} =
    \begin{pmatrix}
        \mu     & \kappa  & \kappa & \dots  & \kappa  \\
        \kappa  & \mu     & \kappa & \dots  & \kappa  \\
        \kappa  & \kappa  & \mu    & \dots  & \kappa  \\
        \vdots  & \vdots  & \vdots & \ddots & \vdots  \\
        \kappa  & \kappa  & \kappa & \dots  & \mu     \\
     \end{pmatrix}.
\end{eqnarray}
We here note that, by following the prescribed steps, the matrix \(S^{-1/2}\) can be effectively generated across varying dimensions and overlap settings \cite{PIELA2014e99}. With the help of \eqref{OperationT}, we obtain the following relations between the orthogonal basis states \(\{\ket{l_0}, \ket{l_1}, \dots, \ket{l_{d-1}}\}\) and the nonorthogonal basis states \(\{\ket{c_0}, \ket{c_1}, \dots, \ket{c_{d-1}}\}\):
\begin{eqnarray}\label{Ddimensional-LSO}
\ket{l_i}=\mu\ket{c_i}+\kappa \sum_{\substack{j=0 \\ (j \neq i)}}^{d-1} \ket{c_j}  \   \   \big(i=0,1,\dots,d-1\big).
\end{eqnarray}
It is easy to check that the basis states \(\{\ket{l_0}, \ket{l_1}, \dots, \ket{l_{d-1}}\}\) are orthonormal indeed. From Eq.~\eqref{Ddimensional-LSO}, with some calculation, nonorthogonal basis states can also be obtained in terms of orthogonal basis states:
\begin{equation}
\ket{c_i}={\sqrt{\lambda_{d-1}\lambda_0}}\bigg[{\Big(\mu+[d-2]\kappa\Big)\ket{l_i}-\kappa\sum_{\substack{j=0 \\ (j \neq i)}}^{d-1}\ket{l_j}}\bigg],
\end{equation}
for \(i=0,1,\dots,d-1\). Thus, we now know the the orthogonal basis states \(\{\ket{l_0}, \ket{l_1}, \dots, \ket{l_{d-1}}\}\) in terms of the nonorthogonal basis states \(\{\ket{c_0}, \ket{c_1}, \dots, \ket{c_{d-1}}\}\), and vice versa. Then, one can find the coefficients \({\bar{\psi}}_i\) such that
\begin{equation}
{\bar{\psi}}_i={\sqrt{\lambda_{d-1}\lambda_0}}\bigg[{\Big(\mu+[d-2]\kappa\Big)\psi_i-\kappa\sum_{\substack{j=0 \\ (j \neq i)}}^{d-1}\psi_j}\bigg],
\end{equation}
for \(i=0,1,\dots,d-1\). Note that, when the overlap matrix \(S\) is equal to the identity matrix \(\dsone\) in Löwdin basis, we have \(\lambda_i=1\), \(\mu=1\), and \(\kappa=0\), and therefore \(\ket{c_i}=\ket{l_i}\) and \({\bar{\psi}}_i=\psi_i\), i.e., two notions, superposition and coherence, coincide when \(s=0\).

Regarding the above standpoint, we can deduce that if a representation of a given quantum state \(\ket{\tau_d}\) is known as a linear combination of one of the two basis states \(\{\ket{c_0}, \dots, \ket{c_{d-1}}\}\) and \(\{\ket{l_0}, \dots, \ket{l_{d-1}}\}\) connected via LSO, then due to the invertibility of the LSO transformation, its representation in the other basis can be easily obtained. Indeed, the LSO method enables efficient conversion between the two bases and allows the original state to be recovered from its representation in either basis. Evidently, the LSO method offers a versatile tool for establishing useful connection(s) between RTC and RTS. In the following, we examine the scenarios where the state \(\ket{\tau_d}\) is regarded as being maximally resourceful.


\section{Main Result: Equivalence of  Maximally Resourceful States}\label{SubSec:MS}

\begin{figure}[t]
	\centering
	\includegraphics[width=.91\columnwidth]{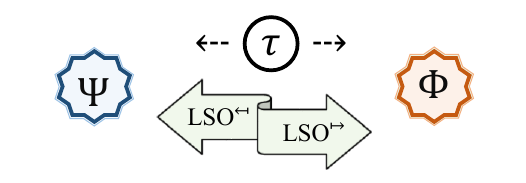}
	\caption{By using the L\"{o}wdin symmetric orthogonalization it is possible to embody the profound connection between the maximally coherent states and states with maximal superposition. Here, \(\Psi\) and \(\Phi\) denote maximal superposition states (i.e., golden states) and maximally coherent states for \(d\)-dimensional systems, respectively. The situation is exactly as follows: With a quantum state \({\tau}\) at hand, if we express \({\tau}\) in terms of Löwdin basis \(\{\ket{l_0},\ket{l_1},\dots,\ket{l_{d-1}}\}\), we have the maximally coherent state \({\Phi}\). But if we express \({\tau}\) in terms of nonorthogonal basis \(\{\ket{c_0},\ket{c_1},\dots,\ket{c_{d-1}}\}\), we have the state with maximal superposition, \({\Psi}\). Essentially, LSO is the only orthogonalization technique that reveals this structure. The detailed analysis with various examples is given in the paper.
}
	\label{Fig2}
\end{figure}

We now proceed to our main result in which we focus our attention on the resource states with maximal superposition. This essential subject was particularly discussed in Ref.~\cite{Senyasa2022Golden},
where the eigensystems of the constructed overlap matrices involve information on the existence of maximal superposition states.
Here, we show that the same results \cite{Senyasa2022Golden} can be obtained forthwith by LSO.
Figure \ref{Fig2} illustrates the way that we apply to realize the form of the states with maximal superposition, i.e., golden states in RTS \cite{Senyasa2022Golden}.


Let us consider the two-dimensional case first. Employing the method of LSO, one obtains the Löwdin basis states \(\{\ket{l_0}, \ket{l_1}\}\) in the basis \(\{\ket{c_0}, \ket{c_1}\}\) with \(\braket{c_0}{c_1}=s\in(-1,1)\), as given in Eq.~\eqref{2D-LSO}.
Let us then assume that one representation of a quantum state \(\ket{\tau_2}\) at hand is given as \(\eta_0\ket{l_0}+\eta_1\ket{l_1}\) in terms of Löwdin basis. With the help of Eq.~\eqref{2D-LSO}, \(\ket{\tau_2}\) can also be written in terms of nonorthogonal basis states such that
\begin{eqnarray}\label{CohToSup}
\ket{\tau_2} \equiv \eta_0\ket{l_0}+\eta_1\ket{l_1} \overset{\text{LSO}^{\mapsfrom}}{\longrightarrow} \zeta_0 \ket{c_0} + \zeta_1 \ket{c_1},
\end{eqnarray}
where \(\zeta_0=\alpha\eta_0+\beta\eta_1\) and \(\zeta_1=\alpha\eta_1+\beta\eta_0\). In other words, if a given quantum state \(\ket{\tau_2}\) could be represented in the basis \(\{\ket{l_0}, \ket{l_1}\}\), then it also has a representation in the basis \(\{\ket{c_0}, \ket{c_1}\}\). Concretely, we have the following
\begin{eqnarray}\label{QubitMaxCoherent1}
\zeta_0 \ket{c_0} + \zeta_1 \ket{c_1} \dashleftarrow \ket{\tau_2} \dashrightarrow
\eta_0\ket{l_0}+\eta_1\ket{l_1}.
\end{eqnarray}
Thus, LSO connects the two states \(\eta_0\ket{l_0}+\eta_1\ket{l_1}\) and \(\zeta_0 \ket{c_0} + \zeta_1 \ket{c_1}\), where the former describes a coherent state and the latter describes a superposition state. Undoubtedly, one of the first questions worth addressing is what form the state \(\zeta_0 \ket{c_0} + \zeta_1 \ket{c_1}\) has if \(\eta_0\ket{l_0}+\eta_1\ket{l_1}\) is taken as the maximally coherent state. To see that, let us consider
\begin{equation}\label{MaxCoherenceD2}
\eta_0 \ket{l_0} + \eta_1\ket{l_1}
=\frac{1}{\sqrt{2}}\left(\ket{l_0}+\ket{l_1}\right),
\end{equation}
which is the maximally coherent state \(\ket{\Phi_{2}}\) for qubit systems. From Eqs.~\eqref{CohToSup} and \eqref{MaxCoherenceD2}, it is easy to get
\begin{eqnarray}\label{MaxSupD2Positive}
\ket{\Phi_{2}} \overset{\text{LSO}^{\mapsfrom}}{\longrightarrow} \ket{\Psi^{+}_{2}}
=\frac{1}{\sqrt{2(1+s)}}(\ket{c_0} + \ket{c_1}).
\end{eqnarray}
The state \(\ket{\Psi^{+}_{2}}\) given in Eq.~\eqref{MaxSupD2Positive} is the maximal superposition state for \(d=2\) with \(\braket{c_0}{c_1} = s \in (-1,0]\)  \cite{Plenio-RTofS,torun2020resource,Senyasa2022Golden}. Note that the state \((\ket{l_0}-\ket{l_1})/\sqrt{2}\) is also the maximally coherent state for the qubit systems. In this case one has
\begin{eqnarray}\label{MaxSupD2Negative}
\frac{1}{\sqrt{2}}\left({\ket{l_0}-\ket{l_1}}\right) \overset{\text{LSO}^{\mapsfrom}}{\longrightarrow}  \ket{\Psi^{-}_{2}}
=\frac{1}{\sqrt{2(1-s)}}(\ket{c_0} - \ket{c_1}).
\end{eqnarray}
The state \(\ket{\Psi^{-}_{2}}\) given in Eq.~\eqref{MaxSupD2Negative} is the maximal superposition state for \(d=2\) with \(s \in [0,1)\) \cite{torun2020resource,Senyasa2022Golden}. In other words, states given in Eqs.~\eqref{MaxSupD2Positive} and \eqref{MaxSupD2Negative} can be
transformed into any state in dimension two with \(s \in (-1,0]\) and \(s \in [0,1)\), respectively, via superposition-free operations. As a consequence, LSO demonstrates that we have
\begin{eqnarray}\label{QubitMaxCoherentGen}
\frac{1}{\sqrt{2(1 \mp {s})}}(\ket{c_0} \mp \ket{c_1}) \dashleftarrow \ket{\tau_2} \dashrightarrow
\frac{1}{\sqrt{2}}\left({\ket{l_0} \mp \ket{l_1}}\right).
\end{eqnarray}
Indeed, the method of LSO is simple and straightforward, and exhibits a distinct feature --- establishes the link between the maximally coherent states and the states with maximal superposition.
We conclude that, in the case of two-dimensional systems, maximally coherent states and states with maximal superposition are equivalent under LSO. This equivalence highlights the structural similarities between coherence and superposition and demonstrates the utility of LSO in analyzing resource states in RTC and RTS.

For the sake of clarity of the discussion given in this section, we now consider the qutrit case. Also, let us examine this for different overlaps setting, so that we can see how we should analyze the same in the higher dimensions. For instance, consider \(\braket{c_0}{c_1}=s\) and \(\braket{c_0}{c_2}=\braket{c_1}{c_2}=-s\). With this setting of overlaps, we first introduce
\begin{eqnarray}
\mu=\frac{1}{3}\left[\frac{1}{\sqrt{1+2s}}+\frac{2}{\sqrt{1-s}}\right], \   \
\kappa=\frac{1}{3}\left[\frac{1}{\sqrt{1+2s}}-\frac{1}{\sqrt{1-s}}\right], \quad
\end{eqnarray}
where \(1+2s\) and \(1-s\) are the eigenvalues of the overlap matrix, latter one being two-fold degenerate.
Then, the method of the symmetric orthogonalization given by Eq.~\eqref{LSO-Elements} provides
\begin{eqnarray}\label{3D-LSO1-Example1}
& \ket{l_0} = \mu \ket{c_0} + \kappa \big(\ket{c_1} - \ket{c_2}\big), \  \
\ket{l_1} = \mu \ket{c_1} + \kappa \big(\ket{c_0} - \ket{c_2}\big), & \nonumber \\
& \ket{l_2} = \mu \ket{c_2} - \kappa \big(\ket{c_0} + \ket{c_1}\big). &
\end{eqnarray}
As a result, we have two basis sets, and the relationship between them is as in Eq.~\eqref{3D-LSO1-Example1}. Now suppose that a quantum state \(\ket{\tau_3}\) is given. The question then is whether the state \(\ket{\tau_3}\) corresponds to a maximally resourceful state both in terms of Löwdin basis \(\{\ket{l_0},\ket{l_1},\ket{l_{2}}\}\) and in terms of nonorthogonal basis \(\{\ket{c_0},\ket{c_1},\ket{c_{2}}\}\). We show that this is indeed the case. Consider, for instance, \(\ket{\tau_3}\) is written as \((\ket{l_0}+\ket{l_1}-\ket{l_2})/\sqrt{3}\) in terms of Löwdin basis, which is one of the maximally coherent state for \(d=3\). Then, substituting Eq.~\eqref{3D-LSO1-Example1} into this state, one obtains \((\ket{c_0} + \ket{c_1} - \ket{c_2})/{\sqrt{3(1+2s)}}\) which is the state with maximal superposition with \(s \in (-\frac{1}{2},0]\) \cite{Senyasa2022Golden}. Thus, we have
\begin{eqnarray}\label{QutritMaxCoherent1}
\frac{\ket{c_0} + \ket{c_1} - \ket{c_2}}{\sqrt{3(1+2s)}} \dashleftarrow \ket{\tau_3} \dashrightarrow
\frac{\ket{l_0}+\ket{l_1}-\ket{l_2}}{\sqrt{3}}. \quad
\end{eqnarray}
In simple terms, LSO defined with respect to the setting \(\braket{c_0}{c_1}=s\) and \(\braket{c_0}{c_2}=\braket{c_1}{c_2}=-s\) with \(s \in (-\frac{1}{2},0]\) connects a maximal coherent state and a state with maximal superposition in dimension \(d=3\) as given in Eq.~\eqref{QutritMaxCoherent1}. In addition, one may ask what happens if the state \(\ket{\tau_3}\) has the form \((\ket{l_0}-\ket{l_1}+\ket{l_2})/\sqrt{3}\) instead of \((\ket{l_0}+\ket{l_1}-\ket{l_2})/\sqrt{3}\) in terms of Löwdin basis, which is also one of the maximally coherent state for qutrit systems. As expected, substituting Eq.~\eqref{3D-LSO1-Example1} into \((\ket{l_0}-\ket{l_1}+\ket{l_2})/\sqrt{3}\), one obtains a superposition state which does not contain maximal superposition. The reason is that we need another setting of overlaps. For instance, let us consider \(\braket{c_0}{c_1}=\braket{c_1}{c_2}=s\) and \(\braket{c_0}{c_2}=-s\). With this setting of overlaps, we first introduce
\begin{eqnarray}
\mu=\frac{1}{3}\left[\frac{1}{\sqrt{1-2s}}+\frac{2}{\sqrt{1+s}}\right], \   \
\kappa=\frac{1}{3}\left[\frac{1}{\sqrt{1-2s}}-\frac{1}{\sqrt{1+s}}\right], \quad \
\end{eqnarray}
where \(1-2s\) and \(1+s\) are the eigenvalues of the overlap matrix, latter one being two-fold degenerate. Then, the symmetric orthogonalization given by Eq.~\eqref{LSO-Elements} provides
\begin{eqnarray}\label{3D-LSO1-Example2}
& \ket{l_0} = \mu \ket{c_0} - \kappa \big(\ket{c_1} - \ket{c_2}\big), \
\ket{l_1} = \mu \ket{c_1} - \kappa \big(\ket{c_0} + \ket{c_2}\big), & \nonumber \\
& \ket{l_2} = \mu \ket{c_2} + \kappa \big(\ket{c_0} - \ket{c_1}\big), &
\end{eqnarray}
Now suppose \(\ket{\tau_3}\) is written as \((\ket{l_0}-\ket{l_1}+\ket{l_2})/\sqrt{3}\) in terms of Löwdin basis. Then, substituting Eq.~\eqref{3D-LSO1-Example2} into this state, one obtains \((\ket{c_0} - \ket{c_1} + \ket{c_2})/{\sqrt{3(1-2s)}}\) which is the state with maximal superposition with \(s \in [0,\frac{1}{2})\) \cite{Senyasa2022Golden}. Thus, we have
\begin{eqnarray}\label{QutritMaxCoherent2}
\frac{\ket{c_0} - \ket{c_1} + \ket{c_2}}{\sqrt{3(1-2s)}} \dashleftarrow \ket{\tau_3} \dashrightarrow
\frac{\ket{l_0}-\ket{l_1}+\ket{l_2}}{\sqrt{3}}, \quad
\end{eqnarray}
where \(\braket{c_0}{c_1}=\braket{c_1}{c_2}=s\) and \(\braket{c_0}{c_2}=-s\) for \(s \in [0,\frac{1}{2})\).
It turns out that, it is possible to bring to light information regarding the states with maximal superposition in dimension \(d\) (see Fig.~~\ref{Fig2}). However, it has to be noted that, each specified overlaps setting links the different forms of maximal states as seen from Eqs.~\eqref{QutritMaxCoherent1} and \eqref{QutritMaxCoherent2}. For instance, we have
\begin{eqnarray}\label{QutritMaxCoherent3}
\frac{-\ket{c_0} + \ket{c_1} + \ket{c_2}}{\sqrt{3(1-2s)}} \dashleftarrow \ket{\tau_3} \dashrightarrow
\frac{-\ket{l_0}+\ket{l_1}+\ket{l_2}}{\sqrt{3}},
\end{eqnarray}
where this connection is recognized for the setting \(\braket{c_0}{c_1}=\braket{c_0}{c_2}=s\) and \(\braket{c_1}{c_2}=-s\) with \(s \in [0,\frac{1}{2})\); or
\begin{eqnarray}\label{QutritMaxCoherent4}
\frac{-\ket{c_0} + \ket{c_1} + \ket{c_2}}{\sqrt{3(1+2s)}} \dashleftarrow \ket{\tau_3} \dashrightarrow
\frac{-\ket{l_0}+\ket{l_1}+\ket{l_2}}{\sqrt{3}},
\end{eqnarray}
where this is recognized for the setting \(\braket{c_0}{c_1}=\braket{c_0}{c_2}=-s\) and \(\braket{c_1}{c_2}=s\) with \(s \in (-\frac{1}{2},0]\). As a result, maximally coherent states and states with maximal superposition for \(d=3\) are equivalent under LSO, akin to the corresponding scenario observed for qubit systems.

As a final example, consider the setting \(\braket{c_i}{c_j}=s \in \mathbb{R}\), where one obtains the Löwdin basis states as given in Eq.~\eqref{Ddimensional-LSO}. Once Löwdin basis states are explicitly defined in terms of nonorthogonal basis, the question is as follows: Given that \(\ket{\tau_d}\) as \(\ket{\Phi_{d}}=({1}/{\sqrt{d}}) \sum_{i=0}^{d-1}\ket{l_i}\), can we identify the form of the state with maximal superposition? Combining Eqs.~\eqref{TheTermMu}, \eqref{TheTermKappa}, and \eqref{Ddimensional-LSO}, and then substituting these into \(\ket{\Phi_{d}}\), we obtain \({(\ket{c_0}+\dots+\ket{c_{d-1}})}/{\sqrt{d(1+[d-1]s)}}\).
Thus, we have
\begin{eqnarray}\label{MaxSupDdimPositive}
\ket{\Psi^{+}_{d}}=\frac{\sum_{i=0}^{d-1}\ket{c_i}}{\sqrt{d(1+[d-1]s)}}
\dashleftarrow \ket{\tau_d} \dashrightarrow \ket{\Phi_{d}}=\frac{1}{\sqrt{d}}\sum_{i=0}^{d-1}\ket{l_i}. \quad
\end{eqnarray}
The state \(\ket{\Psi^{+}_{d}}\) given in Eq.~\eqref{MaxSupDdimPositive} is the maximal superposition state
with \(s \in (\frac{1}{1-d}, 0]\)  \cite{Senyasa2022Golden}. Extensions of the examples similar to the case we explicitly discussed for \(d=2,3\) are possible for \(d\)-dimensional systems. We refer the reader who wants to know more about the states with maximal superposition to Ref.~\cite{Senyasa2022Golden}.

In essence, one of the main advantages of LSO is now manifest: namely, if one knows the form(s) of the maximally coherent states,
then one can surely find out what form the states with maximal superposition are in by using LSO, and vice versa.
In light of the above results, we can conclude that LSO not only preserves the structure and symmetry of the basis vectors
as stated by Eq.~\eqref{LSO-min}, but also serves as a mirror facing the states with maximal coherence (superposition) and reflecting the states with maximal superposition (coherence). We believe that it is very valuable to bring such an alternative and simple solution to the problem, where orthogonalization methods other than LSO do not have such a special feature.


\begin{figure}[t]
\centering
\includegraphics[width=1\columnwidth]{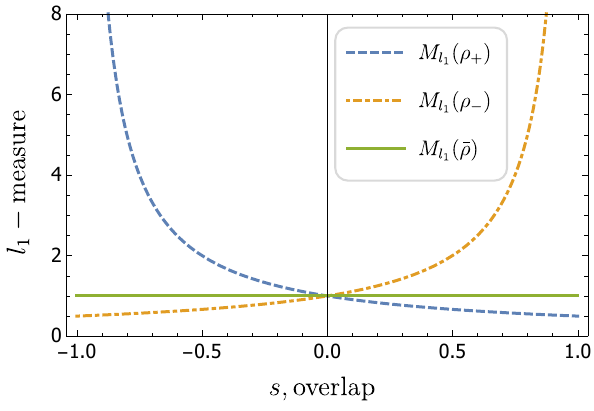}
\caption{Plots of the \(l_1\) measure of the superposition states \(\rho_{+}=\ket{\Psi_2^{+}}\bra{\Psi_2^{+}}\) (the dashed blue line) and \(\rho_{-}=\ket{\Psi_2^{-}}\bra{\Psi_2^{-}}\) (the dash-dotted orange line), and the coherent state \(\bar{\rho}=\ket{\Phi_2}\bra{\Phi_2}\) (the solid green line). Here, \(M_{l_1}(\rho_{+})={1}/{(1+s)}\), \(M_{l_1}(\rho_{-})={1}/{(1-s)}\), and \(M_{l_1}(\bar{\rho})=1\), where \(M_{l_1}(\rho_{+})=M_{l_1}(\rho_{-})=1\) for \(s=0\) as expected.}
\label{Fig3}
\end{figure}

It is clear that the \(l_1\) measure of superposition \cite{Plenio-RTofS} reaches its maximum value for certain superposition states when the overlaps between basis states take on specific values. Namely, for qubit systems, the states \(\ket{\Psi_{2}^{+}}\) and \(\ket{\Psi_{2}^{-}}\) specified in Eqs.~\eqref{MaxSupD2Positive} and \eqref{MaxSupD2Negative} attain their maximum \(l_1\) measure of superposition for values of the overlap \(s \in \mathbb{R}\) within the ranges \((-1,0]\) and \([0,1)\), respectively (see Fig.~\ref{Fig3}). Also note that, as expected, the states \(\ket{\Psi_{2}^{+}}\) and \(\ket{\Psi_{2}^{-}}\) reduce to two unitarily (i.e., incoherent unitary) equivalent forms of the maximal state of coherence \cite{Baumgratz-Coherence} for \(d=2\) in the limit of overlapping between basis states going to zero (i.e., in the orthonormal limit). It goes without saying that these consequences are equally applicable to any system with high dimensions. In particular, an upper bound on the coherence of a superposition of two states in terms of the coherence of the individual states comprising the superposition was presented \cite{Yuwen2019}, utilizing the \(l_1\) measure of coherence. Within this context, it would be useful to explore the implications that emerge from the combined integration of nonorthogonality and LSO.

Note that the above brief discussion has been done using \(l_1\) measure since it is easy to calculate --- other measures such as relative entropy of superposition requires optimization \cite{Plenio-RTofS}, which is particularly difficult to calculate in the nonorthogonal setting. We hope that results related with LSO that connects the theories of coherence and superposition may be useful in investigating effective and practical superposition resource measures. We leave open the capability of LSO for investigating superposition measures, which would be an important direction to consider in the future works on the theory of superposition.


\section{Conclusion and Outlook}\label{Sec:Conc}

To summarize, we have shown that LSO offers compelling advantages in revealing the strong connections between the notions of coherence and superposition within their resource-theoretic formulations. At the heart of our consideration is LSO \cite{Lowdin1950,PIELA2014e99}, which has a unique feature among all existing orthogonalization techniques. Namely, LSO  is distinguished by its geometric approach, which enables it to maintain the original structure and symmetry of nonorthogonal basis states with minimal distortion. Taking cue from this, we have shown that one can obtain the states with maximal superposition (i.e., golden states) from the maximally coherent state by means of LSO. As we proceeded, we were able to completely solve the problem in two and three dimensions, using the methods we have developed along the way. We have obtained consistent results with the comprehensive study recently presented in Ref.~\cite{Senyasa2022Golden}. The method we have devised allows for the realization of maximal superposition states in any dimension. We have also briefly discussed the \(l_1\) measures of the superposition states and the coherent states, where LSO connects them.

Looking forward, there could be various implications of the present results. For instance, practical and efficient strategies were presented to obtain one of a family of maximally coherent states of dimension \(n=2,3,\dots,d\) from any pure state \cite{Torun-CoherenceDistill} and mixed state \cite{Liu2021OptCoherenceDistill} of dimension \(d\). One direction for future work would be to explore whether we can obtain a set of states with maximal superposition of dimension \(n \leq d\) from a given (pure or mixed) superposition state, with the aid of LSO and tasks developed in Refs.~\cite{Torun-CoherenceDistill,Liu2021OptCoherenceDistill}. It would be interesting to see the consequences of these for mixed superposition states. In light of another crucial aspect, namely resource quantifiers, the benefits of LSO can be further explored. For instance, it was shown \cite{Cui2020MaximalValueCohe} that the maximal value condition of coherence measures holds for mixed states if and only if it holds for a special subset of pure states. Analogous investigation employing LSO as an auxiliary tool could be undertaken to explore superposition in a similar vein. Moreover, LSO may provide further motivation to extend the role of coherence as a resource in metrology \cite{DegenQSensing2017,Giovannetti2011} to superposition. We believe that LSO positively impacts our understanding of the links between orthogonal and nonorthogonal settings, and could provide more for general quantum resource theories.


\begin{acknowledgments}	
I acknowledge support from the TÜBİTAK Research Institute for Fundamental Sciences. I would like to thank Onur Pusuluk for valuable remarks and fruitful discussions about L\"{o}wdin symmetric orthogonalization.
\end{acknowledgments}



%

\end{document}